\documentclass[twocolumn]{aa}

\usepackage{graphicx}
\usepackage{txfonts}
\usepackage{xspace}

\begin{document}

\newcommand{\eee}{1ES\,1959+650\xspace}
\newcommand{\mmm}{M\,87\xspace}


  \title{Detection of TeV Gamma-Rays from the BL Lac \eee
    \\
    in its low states and during a major outburst in 2002
    \thanks{Table 4 is only available in electronic form at the CDS via 
      anonymous ftp to cdsarc.u-strasbg.fr (130.79.125.5) or via 
      http://cdsweb.u-strasbg.fr/Abstract.html}
    }

  \author{
    F.~Aharonian\inst{1},
    A.~Akhperjanian\inst{7},
    M.~Beilicke\inst{4},
    K.~Bernl\"ohr\inst{1},
    H.-G.~B\"orst\inst{5},
    H.~Bojahr\inst{6},
    O.~Bolz\inst{1},
    T.~Coarasa\inst{2},
    J.L.~Contreras\inst{3},
    J.~Cortina\inst{10},
    S.~Denninghoff\inst{2},
    M.V.~Fonseca\inst{3},
    M.~Girma\inst{1},
    N.~G\"otting\inst{4},
    G.~Heinzelmann\inst{4},
    G.~Hermann\inst{1},
    A.~Heusler\inst{1},
    W.~Hofmann\inst{1},
    D.~Horns\inst{1},
    I.~Jung\inst{1},
    R.~Kankanyan\inst{1},
    M.~Kestel\inst{2},
    A.~Kohnle\inst{1},
    A.~Konopelko\inst{1},
    H.~Kornmeyer\inst{2},
    D.~Kranich\inst{2},
    H.~Lampeitl\inst{4},
    M.~Lopez\inst{3},
    E.~Lorenz\inst{2},
    F.~Lucarelli\inst{3},
    O.~Mang\inst{5},
    H.~Meyer\inst{6},
    R.~Mirzoyan\inst{2},
    A.~Moralejo\inst{3},
    E.~Ona-Wilhelmi\inst{3},
    M.~Panter\inst{1},
    A.~Plyasheshnikov\inst{1,8},
    G.~P\"uhlhofer\inst{1},
    R.~de\,los\,Reyes\inst{3},
    W.~Rhode\inst{6},
    J.~Ripken\inst{4},
    J.~Robrade\inst{4},
    G.~Rowell\inst{1},
    V.~Sahakian\inst{7},
    M.~Samorski\inst{5},
    M.~Schilling\inst{5},
    M.~Siems\inst{5},
    D.~Sobzynska\inst{2,9},
    W.~Stamm\inst{5},
    M.~Tluczykont\inst{4},
    V.~Vitale\inst{2},
    H.J.~V\"olk\inst{1},
    C.~A.~Wiedner\inst{1},
    W.~Wittek\inst{2}}

  \institute{
    Max-Planck-Institut f\"ur Kernphysik,
    Postfach 103980, D-69029 Heidelberg, Germany
    \and
    Max-Planck-Institut f\"ur Physik, F\"ohringer Ring 6,
    D-80805 M\"unchen, Germany
    \and
    Universidad Complutense, Facultad de Ciencias
    F\'{\i}sicas, Ciudad Universitaria, E-28040 Madrid, Spain
    \and
    Universit\"at Hamburg, Institut f\"ur
    Experimentalphysik, Luruper Chaussee 149,
    D-22761 Hamburg, Germany
    \and
    Universit\"at Kiel, Institut f\"ur Experimentelle und
    Angewandte Physik,
    Leibnizstra{\ss}e 15-19, D-24118 Kiel, Germany
    \and
    Universit\"at Wuppertal, Fachbereich Physik,
    Gau{\ss}str.20, D-42097 Wuppertal, Germany
    \and
    Yerevan Physics Institute, Alikhanian Br. 2, 375036
    Yerevan, Armenia
    \and
    On leave from  
    Altai State University, Dimitrov Street 66, 656099 Barnaul, Russia
    \and
    Home institute: University Lodz, Poland
    \and
    Now at Institut de F\'{\i}sica d'Altes Energies,
    UAB, Edifici Cn, E-08193, Bellaterra (Barcelona), Spain
    }

  \offprints{
    \\
    N.~G\"otting, 
    \email{Niels.Goetting@desy.de}
    }

  \date{Received / Accepted}

  \abstract{
    TeV $\gamma$-rays from the BL Lac object \eee have been
    measured during the years 2000 and 2001 with a significance of
    5.2\,$\sigma$ at a value of 5.3\,\% of the Crab flux
    and in May 2002 during strong outbursts with $> 23\,\sigma$ at a 
    flux level of up to 2.2~Crab, making \eee the TeV Blazar with 
    the third best event statistics. The deep observation
    of 197.4\,h has been performed
    with the HEGRA stereoscopic system of 5~imaging atmospheric 
    Cherenkov telescopes (IACT system). \eee is located at a redshift
    of $z = $0.047, providing an intermediate distance between the
    nearby Blazars Mkn~421 and Mkn~501, and the much more distant object
    H1426+428. This makes \eee an important member of the class of TeV Blazars
    in view of the absorption of TeV photons
    by the diffuse extragalactic background radiation (DEBRA).  
    The differential energy spectrum of \eee during the flares can
    be fitted
    by a power law with a spectral index of
    $2.83 \pm 0.14_{\mbox{\tiny stat}} \pm 0.08_{\mbox{\tiny sys}}$
    or by a power law with an exponential cut-off at
    $(4.2^{+0.8}_{-0.6~{\mbox{\tiny stat}}} \pm 0.9_{\mbox{\tiny sys}})$\,TeV
    and a spectral index of
    $1.83 \pm 0.15_{\mbox{\tiny stat}} \pm 0.08_{\mbox{\tiny sys}}$.
    The low state differential energy spectrum obtained with lower statistics 
    can be described by a pure power law with a
    spectral index of $3.18 \pm 0.17_{\mbox{\tiny stat}} \pm 0.08_{\mbox{\tiny sys}}$.

    }

  \titlerunning{Detection of TeV Gamma-Rays from the BL Lac \eee}
  \authorrunning{Aharonian et al.}
  \maketitle



\section{Introduction}

Active Galactic Nuclei (AGN) are known to be sources of extragalactic
TeV $\gamma$-radiation. Except for a recent report of a significant
TeV excess from the radio galaxy \mmm observed with the HEGRA Cherenkov
telescopes (\cite{hegra_m87}) all TeV AGN detected so far are of the 
BL Lac class. In these objects the very high energy
photons are believed to originate (possibly due to inverse Compton scattering)
in the relativistic jets oriented at small angles to the observer's line of sight.
To these TeV $\gamma$-ray emitters belong the well studied
BL Lac objects Mkn 421 ($z = $ 0.030) 
and Mkn 501 ($z = $ 0.034). 

Recently, the much more distant BL Lac H1426+428 ($z = $ 0.129) 
has been firmly established as an emitter of TeV $\gamma$-radiation 
by the VERITAS (\cite{horan_2002}, \cite{petry_2002}), 
HEGRA (\cite{hegra_1426}, \cite{hegra_1426_2003}), and
CAT (\cite{cat_1426}) collaborations. Possible further TeV detections 
of BL Lac objects have been reported by different collaborations
for 1ES2344+514 ($z =~$0.044) 
(\cite{catanese_1998}), the BL~Lac 
PKS2155-304 ($z =~$0.117) 
(\cite{chadwick_1999}) and the very distant object 3C\,66A 
($z~=~$0.444) 
(\cite{neshpor_1998}), but no confirmation by any other group was 
stated so far.

The object \eee (\cite{elvis_1992})
with a redshift of
$z = $ 0.047 
was classified as a BL~Lac in 1993 using a specially developed
radio/optical/X-ray technique (\cite{schachter_1993}).
A first detection at TeV energies was reported by the Utah Seven 
Telescope Array collaboration for the 1998 observational season 
(\cite{7ta}). An excess with a statistical significance of 
3.9\,$\sigma$ above 600\,GeV was obtained after 57\,h. 
There was no independent confirmation until the year 2001, when the HEGRA
collaboration reported a detection of \eee 
with a significance of $>$~4\,$\sigma$ (\cite{hegra_1es1959_icrc_2001}), 
resp.~$>$ 5\,$\sigma$ with a larger data set 
(\cite{horns_turku_1959}).

In May 2002, \eee underwent a strong TeV outburst, 
detected by the VERITAS (\cite{Whipple_1959_ApJL}),
HEGRA (\cite{horns_turku_1959}, \cite{kranich_turku_1959}) and CAT 
(\cite{cat_1959}) collaborations. Strong variations with flux levels
of up to 3 times the Crab flux have been measured.

In this Letter the results of extensive observations of \eee
using the HEGRA 
IACT system (\cite{hegra_iact_system_1997}) during the years 2000/2001 
and 2002 are reported.
Astrophysical conclusions concerning the nature of this fourth established
TeV Blazar regarding possible influence by the DEBRA absorption are 
briefly discussed.


\section{Observations and results of analysis}

For the following analysis, a total of 18.4\,h of data from the year~2000,
89.4\,h from 2001, and 89.6\,h from 2002 were used taken with the HEGRA 
IACT system. The dates of the HEGRA observation periods are listed in
Table~\ref{hegra_observations}. The mean zenith angle of all observations 
was 38.8$^\circ$, resulting in a mean energy threshold (defined as the 
peak detection rate for $\gamma$-showers)
of 1.4\,TeV for a Crab-like spectrum (\cite{Konopelko_1999}).
\begin{table}[t]
    \caption{Individual periods of \eee observations with the
      HEGRA IACT system. Listed are
      the total observation times and mean zenith angles $\langle \vartheta
      \rangle$. 
      Typically, each night comprises about~1\,h of observation time.}
    \label{hegra_observations}
    \begin{center}
      \begin{tabular}{lc|cc} \hline \hline
        Date                     &     Year   &   Obs. Time   &     $\langle \vartheta \rangle$ \\
                                 &            &   [h]         &     [$^\circ$] \\
        \hline
        July 28 -- August 2      &     2000   &    8.2        &     40.0 \\
        August 28 -- September 6 &     2000   &   10.2        &     40.7 \\
        \hline
        May 18 -- June 1         &     2001   &   13.0        &     37.4 \\
        June 22 -- June 28       &     2001   &   15.5        &     38.1 \\
        July 11 -- July 22       &     2001   &    8.5        &     38.1 \\
        August 9 -- August 25    &     2001   &   17.5        &     37.2 \\
        September 6 -- September 19 &  2001   &   15.0        &     37.3 \\
        October 11 -- October 21 &     2001   &   19.9        &     40.1 \\
        \hline
        May 18 -- May 21         &     2002   &   10.0        &     41.8 \\
        June 3 -- June 18        &     2002   &   34.4        &     38.5 \\
        July 1 -- July 18        &     2002   &   18.7        &     39.5 \\
        July 31 -- August 16     &     2002   &   20.2        &     38.4 \\
        September 3 -- September 11 &  2002   &    6.3        &     38.6 \\
        \hline
        Sum                      &            &  197.4        &     38.8 \\ \hline
      \end{tabular}
    \end{center}
\end{table}
The \eee data were mainly taken with the complete 5-telescope setup of the
IACT system. A few data runs were taken with only 4 telescopes due to
technical reasons. 
Individual runs were accepted, if well-defined quality criteria were 
fulfilled (see, e.\,g.~\cite{hegra_m87}).
%
About 8\,\% of all data were rejected due to this selection.

\begin{table}
    \caption{
      HEGRA IACT system direction reconstruction algorithms and cuts 
      used for the \eee signal search and for the spectral analysis, 
      respectively.
      The resulting numbers for 4 selected subsets are given:
      ``2000/2001'' including all data used for the first 5.2\,$\sigma$
      detection,
      ``May 2002'' only using data from the nights May 18/19 and 
      May~19/20, 2002 (see Figure~\ref{1ES1959_onoff}).
      The ``low state'' resp.~``high state'' data sets are used for
      the spectral analysis (see text and Figure~\ref{1ES1959_spectra}).
      }
    \label{hegra_numbers}
    \begin{center}
      \begin{tabular}{lll} \hline \hline
                                                   &     signal          &   spectrum \\
        \hline
        stereo algorithm                           &     \#3             &   \#3      \\
        number of images per event \quad           &     $\ge$ 3         &   $\ge$ 3  \\
        shape cut on~~{\em mscw}                   &     $<$ 1.1         &   $<$ 1.1      \\
        angular distance cut~~$\Theta^2 $          &     $<$ 0.012 deg$^2$   &  $<$ 0.014 deg$^2$ \\
        $\alpha = A_{\mbox{\tiny ON}} / A_{\mbox{\tiny OFF}}$ & 0.0758   &   0.1429   \\
        \hline \hline
            &  \underline{\em 2000/2001}  &  \underline{``{\em low state}''}\  \\
        $\mbox{T}_{\mbox{\tiny obs}}$              &       93.8\,h       &    149.6\,h  \\
        $N_{\mbox{\tiny ON}}$                      &      354            &    585       \\
        $N_{\mbox{\tiny OFF}}$                     &     3441            &   1960       \\
        $N_{\mbox{\tiny $\gamma$-candidates}}$     &       93 $\pm$ 19   &    305 $\pm$ 25  \\
        significance ($\sigma$)                    &        5.2          &     14.6     \\
        \hline
            &  \underline{\em May 2002}  &  \underline{``{\em high state}''}  \\
        $\mbox{T}_{\mbox{\tiny obs}}$              &      3.6\,h          &      8.5\,h  \\
        $N_{\mbox{\tiny ON}}$                      &    179               &    275   \\
        $N_{\mbox{\tiny OFF}}$                     &    131               &    141   \\
        $N_{\mbox{\tiny $\gamma$-candidates}}$     &    169 $\pm$ 13      &    255 $\pm$ 17 \\
        significance ($\sigma$)                    &     23.4             &     25.5 \\ \hline
      \end{tabular}
    \end{center}
\end{table}
All observations of \eee were carried out in the so-called 
{\em wobble} mode 
allowing for simultaneous estimation of the background (``OFF'') rate 
induced by charged cosmic rays (\cite{Aharonian_1997}). This
analysis uses for the signal search the so-called ring segment background 
model as explained in \cite{hegra_m87} providing a small ratio of ON to 
OFF-source solid angle areas
$\alpha = A_{\mbox{\tiny ON}}/A_{\mbox{\tiny OFF}}$
(resulting in improved background statistics).
For the spectral analysis a set of 7 independent background regions is 
used (\cite{hegra_cas-a}).

The general shower reconstruction and the event selection cuts for
the image analysis have been described in previous publications 
(e.\,g.~\cite{Aharonian_1999a}). 
For the signal search, the stereo air shower direction
reconstruction algorithm~\#3 (\cite{Hofmann_1999}) and a tight shape cut 
(parameter {\em mscw} $ < 1.1$) (\cite{Konopelko_1999}) for an effective
$\gamma$-hadron separation have been applied. Because of the relatively
large zenith angles more usable shower images per event are recorded
compared to observations close to the zenith.
Therefore, a minimum number of 3 images per event is required for the 
whole reconstruction chain rejecting all 2-telescope events, which only 
provide a poor resolution.
The optimum angular cut was derived using $\gamma$-ray
events from the Crab nebula on the basis of a nearly contemporaneous data set
at similar zenith angles.
The procedure of spectral evaluation -- with an energy resolution of 
$\le 12$\,\% for a single event~-- is discussed in
\cite{hegra_1426_2003}.
The energy spectrum is derived from a background subtracted photon count
spectrum using an effective area on an event by event basis,
depending upon the number of active telescopes, the reconstructed 
energy, the zenith angle and the assumed spectral shape. 
In order to take into account the slightly varying detector performance,
the effective areas are determined on a monthly basis 
along with the mirror reflectivities, photocathode efficiencies
and conversion factors from digitized 
photomultiplier signals to photoelectrons (\cite{hegra_performance_2003}).

\begin{figure}[t!]
  \centering
  \includegraphics[width=8.5cm]{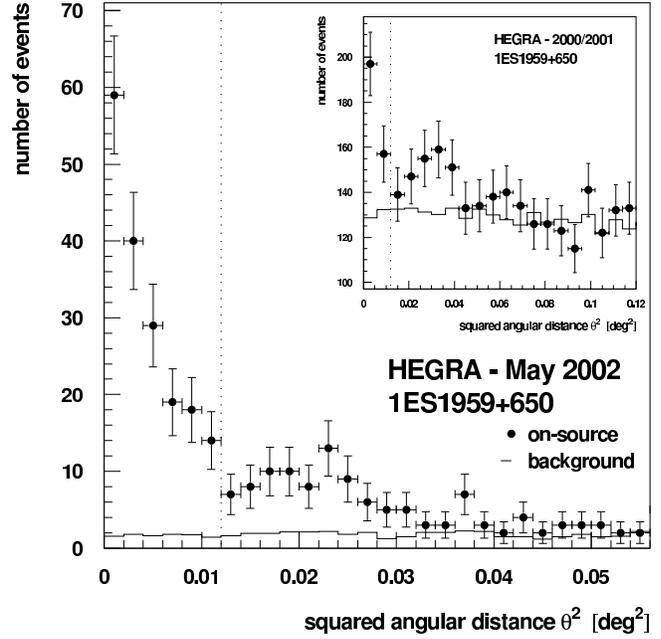}
  \caption{
    Number of events vs.~squared angular distance $\Theta^2$ to the position of
    \eee, observed on May 18/19 and 19/20, 2002, with the HEGRA IACT system. 
    The dots show the ON-source events, the histogram
    gives the background estimate determined from a ring segment
    as explained in the text. The statistical error for
    the background estimate is much smaller than the error of the 
    ON-source distribution. 
    Upper right: Similar distributions for the year~2000/2001 observations
    yielding a significance of 5.2\,$\sigma$.
    }
  \label{1ES1959_onoff}
\end{figure}

\begin{figure}[t!]
  \centering
  \includegraphics[width=8.5cm]{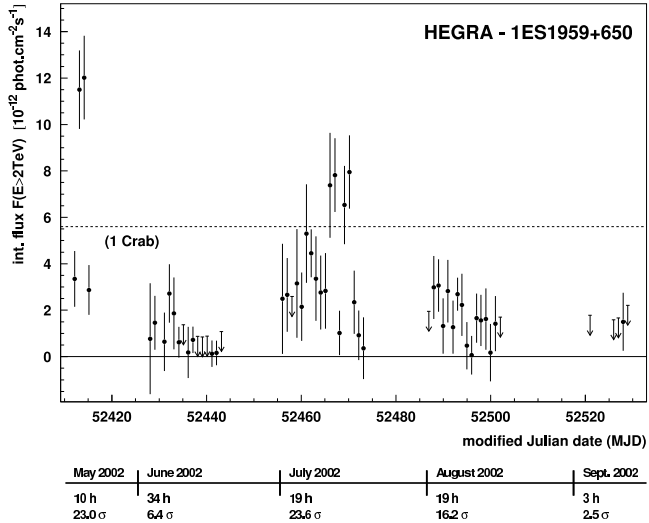}
  \caption{
    Diurnal integral flux values resp.~upper limits
    (99\,\%~c.\,l.) observed above 
    2\,TeV vs.~the modified Julian date (MJD) for the year~2002 
    HEGRA IACT system data. Strong flux 
    variations are visible in May and July. The level of the flux of the
    Crab nebula is indicated by the horizontal dashed line. The time gaps
    between the individual observation periods are due to the full moon
    periods when no observations were possible.
    }
  \label{1ES1959_lightcurve}
\end{figure}

The relevant direction reconstruction algorithms, event selection cuts, 
resulting event numbers, and significances for the individual HEGRA 
\eee data sets used for the signal search and the spectral 
analysis are summarized in Table~\ref{hegra_numbers}.
Figure~\ref{1ES1959_onoff} shows the event distribution
for the ON-source and the OFF-source regions from the May 2002 flare and
the year 2000/2001 quiescent state data sets, respectively, as a function of the squared
angular distance to the source position.
The statistical significance of the excess from the direction 
of \eee amounts to 5.2\,$\sigma$ in the year 2000/2001
observations, calculated using formula 17 in \cite{Li_Ma}.
The evidence for TeV $\gamma$-rays from \eee during this quiescent state
at a value of $(5.3 \pm 1.1_{\mbox{\tiny stat}})$\,\% of the flux of the 
Crab nebula (\cite{hegra_crab_spectrum})
already confirms the tentative detection of \eee by the Utah Seven
Telescope Array. The HEGRA detection during the 
strong outburst in May~2002 is obvious at a very high significance of
greater than 23\,$\sigma$.
%
%
%
The diurnal integral flux levels above 2\,TeV
observed in the year~2002 are shown in Figure~\ref{1ES1959_lightcurve}
(Table 4 containing the complete lightcurve for the years 2000 - 2002 is available at the
CDS, see note to the title). 
The strong outburst in May~2002 is followed by a post-flare low state in June.
During July, the source has exhibited more activity again followed
by a period of diminishing TeV flux in August and September 2002.

The spectral investigation of the \eee ``high state'' was performed using
the data of the 6 nights in May and July 2002
with the object's integral flux being larger than 1 Crab above 2\,TeV
(see Table~\ref{1ES1959_energy_table}).
Within the limited statistics, the single nights of these 8.5\,h of 
observations show the same spectral shape.
A fit of a pure power law  
%
d$N / \mbox{d}E = N_0 \times (E/\mbox{1\,TeV})^{-\alpha}$ 
to the differential energy spectrum results in
$N_0 = (7.4 \pm 1.3_{\mbox{\tiny stat}} \pm 0.9_{\mbox{\tiny sys}}) \times
10^{-11}$\,phot.\,cm$^{-2}$\,s$^{-1}$TeV$^{-1}$
and
$\alpha = 2.83 \pm 0.14_{\mbox{\tiny stat}} \pm 0.08_{\mbox{\tiny sys}}$ with
$\chi^2_{\mbox{\tiny red.}}(\mbox{d.o.f.}) = 1.9\,(6)$.
A fit of a power law with an exponential cut-off
%
d$N / \mbox{d}E = N_0 \times (E/\mbox{1\,TeV})^{-\alpha} \times \exp(-E/E_c)$
with
$N_0 = (5.6 \pm 0.9_{\mbox{\tiny stat}} \pm 0.7_{\mbox{\tiny sys}}) \times
10^{-11}$\,phot.\,cm$^{-2}$\,s$^{-1}$TeV$^{-1}$,
$\alpha~=~1.83 \pm 0.15_{\mbox{\tiny stat}} \pm 0.08_{\mbox{\tiny sys}}$,
$E_c = (4.2^{+0.8}_{-0.6~{\mbox{\tiny stat}}} \pm 0.9_{\mbox{\tiny sys}})$\,TeV,
$\chi^2_{\mbox{\tiny red.}}(\mbox{d.o.f.}) = 1.7\,(5)$,
is also an adequate description of these data
(see Fig.~\ref{1ES1959_spectra}).

Figure \ref{1ES1959_spectra} also shows the spectral energy distribution
of the \eee ``low state''.
Nearly 150\,h of observations during nights with \eee emitting 
TeV $\gamma$-rays at a flux level below 0.5 Crab have been used for this 
purpose. The spectrum can be described by a pure power law with
%
$N_0 = (7.8 \pm 1.5_{\mbox{\tiny stat}} \pm 1.0_{\mbox{\tiny sys}}) \times
10^{-12}$\,phot.\,cm$^{-2}$\,s$^{-1}$TeV$^{-1}$
and
$\alpha~=~3.18 \pm 0.17_{\mbox{\tiny stat}} \pm 0.08_{\mbox{\tiny sys}}$
with
$\chi^2_{\mbox{\tiny red.}}(\mbox{d.o.f.}) = 0.22\,(3)$.
%
%
A power law with an exponential cut-off and a spectral index fixed
at $\alpha = 1.8$ fits the data with
%
%
$N_0 = (6.0 \pm 1.4_{\mbox{\tiny stat}} \pm 0.8_{\mbox{\tiny sys}}) \times
10^{-12}$\,phot.\,cm$^{-2}$\,s$^{-1}$TeV$^{-1}$
and
%
$E_c = (2.7^{+0.6}_{-0.4~{\mbox{\tiny stat}}} \pm 0.6_{\mbox{\tiny sys}})$\,TeV
with
%
$\chi^2_{\mbox{\tiny red.}}(\mbox{d.o.f.}) = 0.65\,(3)$.

\begin{table}[t]
  \caption
  {
    HEGRA IACT system event statistics and values of the differential
    photon flux of \eee in the high state for the individual energy bins as shown in 
    Figure~\ref{1ES1959_spectra}.
    }
  \label{1ES1959_energy_table}
  \begin{center}
    \begin{tabular}{rrrrr} \hline \hline
      Energy  &  $N_{\mbox{\tiny ON}}$  &  $N_{\mbox{\tiny OFF}}$  &  Sign.       &  \multicolumn{1}{c}{diff.~Flux}  \\
      (TeV)   &                         &                          &  $(\sigma)$  &  \multicolumn{1}{c}{(phot.\,cm$^{-2}$\,s$^{-1}$\,TeV$^{-1}$)}  \\ \hline
       1.52 &  42  &  21  &  10.0  &  $(2.5 \pm 0.5) \times 10^{-11}$  \\
       2.02 &  44  &  43  &   8.6  &  $(8.5 \pm 1.6) \times 10^{-12}$  \\
       2.68 &  44  &  25  &  10.0  &  $(4.0 \pm 0.7) \times 10^{-12}$   \\
       3.55 &  54  &  12  &  12.9  &  $(3.0 \pm 0.4) \times 10^{-12}$  \\
       4.70 &  34  &   9  &  10.0  &  $(1.3 \pm 0.3) \times 10^{-12}$  \\
       6.23 &  16  &   3  &   7.1  &  $(4.5 \pm 1.2) \times 10^{-13}$  \\
       8.25 &   6  &   4  &   3.5  &  $(1.1 \pm 0.5) \times 10^{-13}$  \\
      10.94 &   5  &   1  &   4.0  &  $(0.7 \pm 0.3) \times 10^{-13}$  \\
      \hline
    \end{tabular}
  \end{center}
\end{table}
%


\section{Summary and Conclusions}

The BL Lac object \eee has been observed with the HEGRA IACT system 
for a total of 197.4\,h. The object has been detected at TeV energies 
during its low flux state in the years~2000 and 2001 at a value of 
$(5.3 \pm 1.1_{\mbox{\tiny stat}})$\,\% of the Crab flux and during 
flaring states in 2002, reaching a flux level as high as 2.2~Crab. 
Extensive HEGRA observations in 2002 show strong changes in the absolute 
flux level.

The time averaged energy spectrum of the high state observations 
is well described by a pure power law with spectral index
$\alpha = 2.83 \pm 0.14_{\mbox{\tiny stat}} \pm 0.08_{\mbox{\tiny sys}}$ 
or by a power law with an exponential cut-off
at~$E_c = (4.2^{+0.8}_{-0.6~{\mbox{\tiny stat}}} \pm 0.9_{\mbox{\tiny sys}})$\,TeV.
Within the synchrotron self-Compton (SSC) model the 
TeV $\gamma$-ray emission of a BL Lac type object corresponds to the 
Inverse Compton (IC) component.
\cite{costamante_2002} have predicted the IC flux above 1\,TeV in a 
flaring state of \eee to be $1.74 \times 10^{-11}$\,phot.\,cm$^{-2}$\,s$^{-1}$
using a phenomenological parametrization of the spectral energy distribution
by \cite{fossati_1998}. 
Taking into account the strong variations of \eee in~2002 
the observed flux values can be easily accommodated.

\begin{figure}
  \centering
  \includegraphics[width=8.5cm]{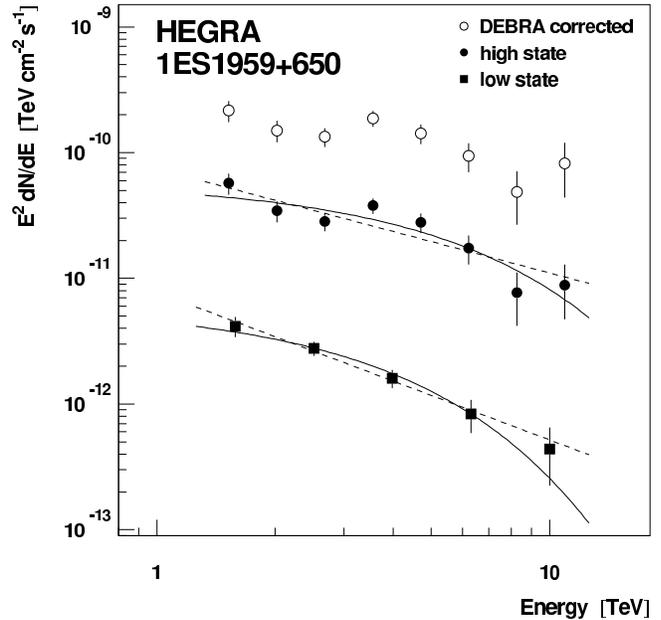}
  \caption
  {
    Time averaged spectral energy distributions (SED) of \eee as measured
    with the HEGRA IACT system. The filled
    circles show the spectrum during 6 high state nights in 2002,
    while the filled squares
    show the SED for the combination of all low state nights ($<$ 0.5 Crab).
    The results of fits of a power law with an exponential cut-off (solid
    lines) resp.~a pure power law (dashed lines) are indicated.
    The open circles show the high state spectrum corrected for the DEBRA 
    absorption (see text).
    }
  \label{1ES1959_spectra}
\end{figure}

Due to the absorption of TeV photons by the pair production process
$\gamma_{\mbox{\tiny TeV}} + \gamma_{\mbox{\tiny DEBRA}} \longrightarrow e^+ + e^-$ 
with the DEBRA the observed spectrum differs from the source spectrum.
In the range from 1 to several TeV however, the spectral shape remains 
nearly unchanged 
since the optical depth in this energy range is rather constant for most of 
the DEBRA candidate spectra in the relevant wavelength range of 
1 to 10\,$\mu$m (\cite{Aharonian_ICRC_2001}).
%
%
Using a ``model of choice'' prediction of the DEBRA spectral energy density
(\cite{Aharonian_ICRC_2001}), the attenuation coefficients for the HEGRA
\eee high state data points have been calculated and applied to 
unfold the measured spectrum as shown in Figure \ref{1ES1959_spectra}. The 
cut-off energy is only marginally shifted 
to~$(4.5^{+1.2}_{-0.8~{\mbox{\tiny stat}}} \pm 0.9_{\mbox{\tiny sys}})$\,TeV.
%
The low event statistics at the highest energies observed around 10\,TeV and
above (Tab.~\ref{1ES1959_energy_table}) do
not allow one to see the strong deformation of the spectral shape,
which is expected from the energy dependence of the attenuation 
coefficients (e.\,g.~see Figure~9 in \cite{Aharonian_ICRC_2001}).

The fact that \eee 
is about~1.5~times more distant than the nearby
TeV Blazars Mkn 421 and Mkn 501 and bridges the distance gap to H1426+428 
makes it a very important object for future DEBRA probing
in the optical to near-infrared region. 
%
%
%
%


\begin{acknowledgements}

  The support of the German Federal Ministry for Research and Technology BMBF and
  of the Spanish Research Council CICYT is gratefully acknowledged. 
  G.~Rowell acknowledges receipt of a von Humboldt fellowship.
  We thank the Instituto de Astrof\'{\i}sica de Canarias (IAC)
  for the use of the HEGRA site at the Observatorio del Roque de los 
  Muchachos (ORM) and for supplying excellent working conditions on La
  Palma.

\end{acknowledgements}


\end{document}